\documentclass[twoside]{article}
\usepackage{graphicx}
\usepackage{fancyhdr}
\usepackage{multicol}
\usepackage{styl-ap}

\def\kms{km$\,$s$^{-\!1}$} 
\def\vsi{$v\: \sin i$}

\def\t{{\scriptsize$\times$}}

\begin{document}
\title{Rotation of the Mass Donors in High-mass X-ray Binaries and Symbiotic Stars}
\author{Kiril Stoyanov\work{1} \& Radoslav Zamanov\work{1}}
\workplace{Institute of Astronomy and National Astronomical Observatory, Bulgarian Academy of Sciences, 
       72 Tsarigradsko Shousse Blvd., 1784 Sofia, Bulgaria
}

\mainauthor{kstoyanov@astro.bas.bg}
\maketitle

\begin{abstract}%
Our aim is to investigate the tidal interaction in 
High-mass X-ray Binaries and Symbiotic stars in order to determine in which objects the rotation of the mass donors 
is synchronized or pseudosynchronized with the orbital motion of the compact companion. We find that the Be/X-ray binaries are not synchronized and the orbital periods of the systems are greater than the rotational periods of the mass donors.
The giant and supergiant High-mass X-ray binaries and symbiotic stars are close to synchronization. 
We  compare the rotation of mass donors in symbiotics  with the
projected rotational velocities of field giants and find that the M giants in S-type symbiotics rotate on average 1.5 times faster than the field M giants. We find that the projected rotational velocity of the red giant in symbiotic star MWC~560 is \vsi $= 8.2 \pm 1.5$~\kms, 
and estimate its rotational period to be  $P_{\rm rot}$ = 144 - 306~days. 
Using the theoretical predictions of tidal interaction and pseudosynchronization, 
we estimate the orbital eccentricity  $e=0.68-0.82$.
\end{abstract}


\begin{multicols}{2}
\section{Introduction}
A High-mass X-ray binary system consists of a compact object (a neutron star or a black
hole) accreting material from an O or B companion star. 
They are divided into Be/X-ray binaries (main-sequence star as a companion) and Giant/Supergiant X-ray binaries (giant or supergiant star as a companion). Accretion of matter is different for both types of X-ray binaries.                                                   In the Be/X-ray binaries, the compact object crosses the circumstellar disc and accretes matter from that disk. 
In the giant/supergiant X-ray binaries, the mass donor ejects a slow and dense wind radially outflowing from the equator and the
compact object directly accretes the stellar wind through Bondi-Hoyle-Lyttleton accretion. 

Symbiotic stars are interacting binaries, consisting of an evolved giant (either a
normal red giant in S-types symbiotics or a Mira-type variable in D-types symbiotics) transferring mass to a hot and luminous white dwarf or neutron star. The symbiotic stars are surrounded by a rich and luminous nebula resulting from the presence of both
an evolved giant with a heavy mass-loss and of a hot companion abundant in ionizing photons and often emanating its own wind.

\section{Synchronization and Pseudosynchronization}
In a binary with a circular orbit the rotational period of the primary, P$_{rot}$, 
reaches an equilibrium value at the orbital period, $P_{orb}=P_{rot}$.  
In other words the synchronous rotation (synchronization) means that 
the rotational period is equal to the orbital period. 
In a binary with an eccentric orbit,
the corresponding equilibrium is reached at a value of $P_{rot}$ which is
less than $P_{orb}$, the amount less being a function solely of the orbital
eccentricity $e$. In practice, in a binary with an eccentric orbit the 
tidal force acts to synchronize the rotation of the mass donor with 
the motion of the compact object at the periastron - 
pseudosynchronous rotation.
To calculate the period of pseudosynchronization, P$_{ps}$, we use (Hut 1981):
\begin{equation}
P_{ps} = \frac{(1+3e^2+\frac{3}{8}e^4)(1-e^2)^\frac{3}{2}}{1+\frac{15}{2}e^2+
\frac{45}{8}e^4+\frac{5}{16}e^6} P_{orb}.
\label{Eq-ps}
\end{equation}

\subsection{Stars with  radiative envelopes}
Following Hurley, Tout \& Pols (2002) the circularization 
timescale for stars with  radiative envelopes is:
\begin{equation}
\frac{1}{{\tau}_{circ}}= 
\frac{21}{2} \left( \frac{G M_1}{R_1^3} \right)^\frac{1}{2} 
q_2 \left(1 + q_2 \right)^\frac{11}{6} E_2 \left(\frac{R_1}{a} \right)^\frac{21}{2},  
\label{circ_re}
\end{equation}
where M$_1$ and R$_1$ are the mass and the radius of the primary respectively, 
q$_2$ is the mass ratio M$_2/$M$_1$, and $a$ is the semi-major axis.
The second-order tidal coefficient 
E$_2$ = 1.592 $\times 10^{-9} M_1^{2.84}$.  

The synchronization time scale is given as, 
\begin{equation}
\tau_{sync} = K \tau_{circ},
\label{sync_re}
\end{equation}
where K  is: 
\begin{equation}
K \approx \frac{0.015}{r_g} \, \frac{1 + q_2}{q_2} \left( \frac{R_1}{a} \right)^2.  
\end{equation}
For the gyration radius of the primary $r_g$
we adopt $r_g\approx0.16$ for giants, and $r_g\approx0.25$ for main sequence stars
(Claret \& Gimenez, 1989).

\subsection{Stars with convective envelopes}
Following Hurley, Tout \& Pols (2002) the synchronization timescale for stars with convective
envelopes is:
\begin{equation}
 \tau_{\rm syn} \approx 800 \left( \frac{ M_{\rm 1}  R_{\rm 1}}{ L_{\rm 1}}\right)^{1/3} 
 \frac{M_{\rm 1}^2 (\frac{M_{\rm 1}}{M_{\rm 2}} + 1)^2}{R_{\rm 1}^6} P_{\rm orb}^4\ \;   {\rm yr},
\label{sync}
\end{equation}
where $L_{\rm 1}$ is the luminosity of the giant. 

The circularization time scale is:
\begin{equation}
 \frac{1}{{\tau}_{\rm circ}} = \frac{21}{2} \left(\frac{k}{T}\right) q{_2} (1+q{_2}) 
 \left(\frac{R_{\rm 1}}{a}\right)^8 .
\label{circ}
\end{equation}
In  Eq.~\ref{circ}, $(k/T)$ is:
\begin{equation}
\left(\frac{k}{T}\right) = \frac{2}{21} \frac{f_{\rm conv}}{{\tau}_{\rm conv}} \frac{M_{\rm env}}{M_{\rm 1}}  
\;  {\rm yr}^{-1} ,
\label{KT}
\end{equation}
where $R_{\rm env}$ is the depth of the convective envelope, $M_{\rm env}$ is the envelope's mass, and
\begin{equation}
{\tau}_{\rm conv} = 0.4311 \left(\frac{M_{\rm env}R_{\rm env}(R_{\rm 1}-\frac{1}{2}R_{\rm env})}{3 L_{\rm 1}}\right) ^{\frac{1}{3}}  {\rm yr}
\label{tau}
\end{equation}
is the eddy turnover time scale (the time scale on which the largest convective cells turnover). 
The numerical factor $f_{\rm conv}$ is
\begin{equation}
{f}_{\rm conv} = {\rm min} \left[1, \left( \frac{P_{\rm tid}}{2{\tau}_{\rm conv}} \right) ^2 \right],
\label{fconv}
\end{equation}
where $P_{\rm tid}$ is the tidal pumping time scale given by
\begin{equation}
\frac{1}{P_{\rm tid}} = \left|\frac{1}{P_{\rm orb}} - \frac{1}{P_{\rm rot}}\right|.
\label{ptid}
\end{equation}

The pseudosynchronization timescale is $\tau_{\rm ps}$ = (7/3($\alpha$ - 3)) $\tau_{\rm circ}$,
where $\alpha$ is a dimensionless quantity, representing the ratio of the orbital and rotational angular
momentum:

\begin{equation}
\alpha = \frac {q_2}{1+q_2} \frac {1}{r_{\rm g}^{2}} \left( \frac {a}{R_{\rm 1}} \right) ^2 .
\end{equation}
For a red giant we adopt $r_{\rm g} \approx 0.3$ (Claret, 2007).

In all equations, the masses, the radii and the lumunosities are in solar units.

\section{High-mass X-ray Binaries}
\vspace{-8mm}
The orbital and stellar parameters of 13 High-mass X-ray binaries are given in Table~1 
and Table~2 in Stoyanov \& Zamanov (2009). We add 2 more objects - 4U~2206+54 and MWC~148. The
orbital and stellar parameters are taken from Rib{\'o} et. al. (2006) and Casares et al. (2012) respectively.

Using Eq.\ref{circ_re} and Eq.\ref{sync_re} we estimate the circularization and
synchronization timescales. The results are given in Table~\ref{t-times}. 
The lifetime of a star on the main sequence can be estimated as
$\tau_{MS} = 10^{10} ({M_\odot}/{M})^{2.5} \; \; {\rm years}$ (Hansen \& Kawaler, 1994).  
Comparing these  lifetimes with $\tau_{sync}$ 
from Table~\ref{t-times}, 
we see that among the Be/X-ray binaries only for LSI+61$^0$303 is
$\tau_{sync} \sim \tau_{MS}$.
This is the only Be/X-ray binary for which we can expect
considerable changes of the rotation  of the primary
during the lifetime of the Be star. 

The lifetime of the giant is comparable or longer then 
$\tau_{circ}$ \ and  \  $\tau_{sync}$  \  for the giant/supergiant systems
with short orbital periods. The exceptions are V725~Tau and BP~Cru, for
which $\tau_{sync}$ and $\tau_{circ}$ are longer than the lifetime of the giant/supergiant stage. 

On Fig.1 in Stoyanov \& Zamanov (2009) is plotted P$_{rot}$ versus P$_{ps}$. 
The giant/supergiant systems are located close to the line 
P$_{ps} = P_{rot}$, 
while those with mass donors from spectral class V are far away from the equilibrium.

In the Be/X-ray systems BQ~Cam, V635 Cas, V725 Tau and 4U~2206+54, the tidal force spinning down the donor star.
For the system LSI+61$^0$303, the rotation of the mass donor 
is close to pseudosynchronization.
This is the only Be/X-ray binary in which  $\tau_{sync}$ is comparable with the 
life-time of the binary. In the binaries X~Per and MWC~148, the neutron star is far away from the Be star
and the tidal force is weak. 

Giant and supergiant systems are close to (pseudo)synchronization.
In these binaries 
the rotation of the mass donors is influenced by the presence 
of the compact object.
In LMC X-4 and Cen X-3, the mass donors 
are synchronized and the orbits are circularized. 
With respect to the rotation of the mass donor, V725~Tau 
is similar to the Be/X-ray binaries. 
Cyg X-1 is synchronized and almost circularized. 
V830~Cen is pseudosynchronized but not circularized yet.
The systems LSI$+65^0010$ and Vela X-1 are close to 
pseudosynchronization and the tidal force accelerates 
the rotation of the mass donors. 
In the case of SMC X-1, the tidal force acts as a decelerator 
of the rotation of the mass donor.
In BP~Cru, a gas stream from the mass donor  exists, 
probably resulting from the strong tidal force and spin-up of the mass donor (Leahy \& Kostka, 2008).   

\begin{mytable}
\caption{Calculated time scales. Given here are  
  the name of the object,  
  synchronization time scale, 
  circularization time scale, and the lifetime.}
\label{t-times}
{\begin{tabular}{lllc}
\hline
object   &    $\tau_{sync}$ [yr] & $\tau_{circ}$ [yr] & lifetime [yr]  \\
\hline
\multicolumn{2}{l}{{\bf Be/X-ray binaries}}  \\
LSI+61$^0$303*&   2.8\t10$^7$    & 2.4\t10$^8$    &5.6\t10$^7$ \\
X~Per         &   6.2\t10$^{17}$ & 1.8\t10$^{21}$ &1.1\t10$^7$ \\
BQ~Cam        &   3.5\t10$^{11}$ & 7.6\t10$^{13}$ &3.9\t10$^6$ \\
V635~Cas      &   1.4\t10$^{11}$ & 9.5\t10$^{12}$ &7.3\t10$^6$ \\
4U~2206+54    &   4.9\t10$^9$	 & 3.7\t10$^{11}$ &7.3\t10$^7$ \\
MWC~148       &   1.2\t10$^{17}$ & 5.1\t10$^{20}$ &9.8\t10$^7$ \\
\multicolumn{2}{l}{{\bf Giant systems}}  \\
V725~Tau      &   2.8\t10$^{12}$ & 8.0\t10$^{14}$ &4\t$10^5$   \\
LMC X-4       &   4.5\t10$^2$    & 7.7\t10$^2$    &1\t$10^6$   \\
Cen~X-3       &   2.3\t10$^3$    & 4.2\t10$^3$    &5\t$10^5$   \\
\multicolumn{2}{l}{{\bf Supergiant systems}} \\
V830~Cen      &   7.5\t10$^3$ & 1.4\t10$^4$ & 1\t10$^6$   \\
LSI+65$^0$010 &   1.3\t10$^4$ & 3.9\t10$^4$ & 1\t10$^6$  \\
Vela~X-1      &   1.0\t10$^4$ & 2.8\t10$^4$ & 3.9\t10$^5$ 	   \\
SMC X-1       &   3.3\t10$^4$ & 8.2\t10$^4$ & 8.8\t10$^5$    \\		      
BP Cru        &   1.8\t10$^6$ & 8.8\t10$^6$ & 8\t10$^4$     \\
Cyg X-1	      &     $<1$      & $<1$       & 1\t10$^5$    \\
\hline
\end{tabular}}
*assuming neutron star as a secondary component.
\end{mytable}

\vspace{-10mm}
\section{S-type symbiotic stars}
43 symbiotic stars have been observed with FEROS spectrograph  at the 2.2m ESO
telescope of the La Silla Observatory (Zamanov et al. 2007). 
The data for the rotation of 55 field red giants are taken from the 
literature. M giants in S-type symbiotics rotate faster than 
the field M giants. 
Histograms of the available \vsi\  data  for the red giants
are plotted in Fig.2 in Zamanov \& Stoyanov (2012).  
For the field  M0III-M6III giants we calculate a
mean \vsi$=$5.0~\kms,  median \vsi$=$4.3~\kms, and
standard deviation of the mean $\sigma=4.0$~\kms.
For the M0III-M6III giants in symbiotics,
we get a mean \vsi$=7.8$~\kms,   median \vsi$=$8.0~\kms,
and standard deviation of the mean $\sigma=$2.1~\kms.

There are 5 objects in our sample that deviate from the synchronization. 
These objects are 
RS~Oph, MWC~560, CH~Cyg, CD-43$^\circ$14304 and Z~And. 
In three of them  collimated jets are detected: 
Z~And   (Skopal et al. 2009); 
CH~Cyg  (Crocker et al. 2002), MWC~560 (Tomov et al. 1990). 
Additionally to the jets,  ejection of blobs are detected from  RS~Oph and CH~Cyg
(Iijima et al. 1994). This confirms  the  suggestions that in the jet-ejecting symbiotics 
the mass donors  rotate faster than the orbital periods.
Probably there is a link between the jets  
and the mass donor rotation.

On Fig.\ref{stoyanov-fig1} are plotted together the High-mass X-ray binaries and the S-type symbiotic stars.
It shows that none of the objects in our sample is above the line of synchronization.

\begin{myfigure}
\centerline{\resizebox{90mm}{!}{\includegraphics{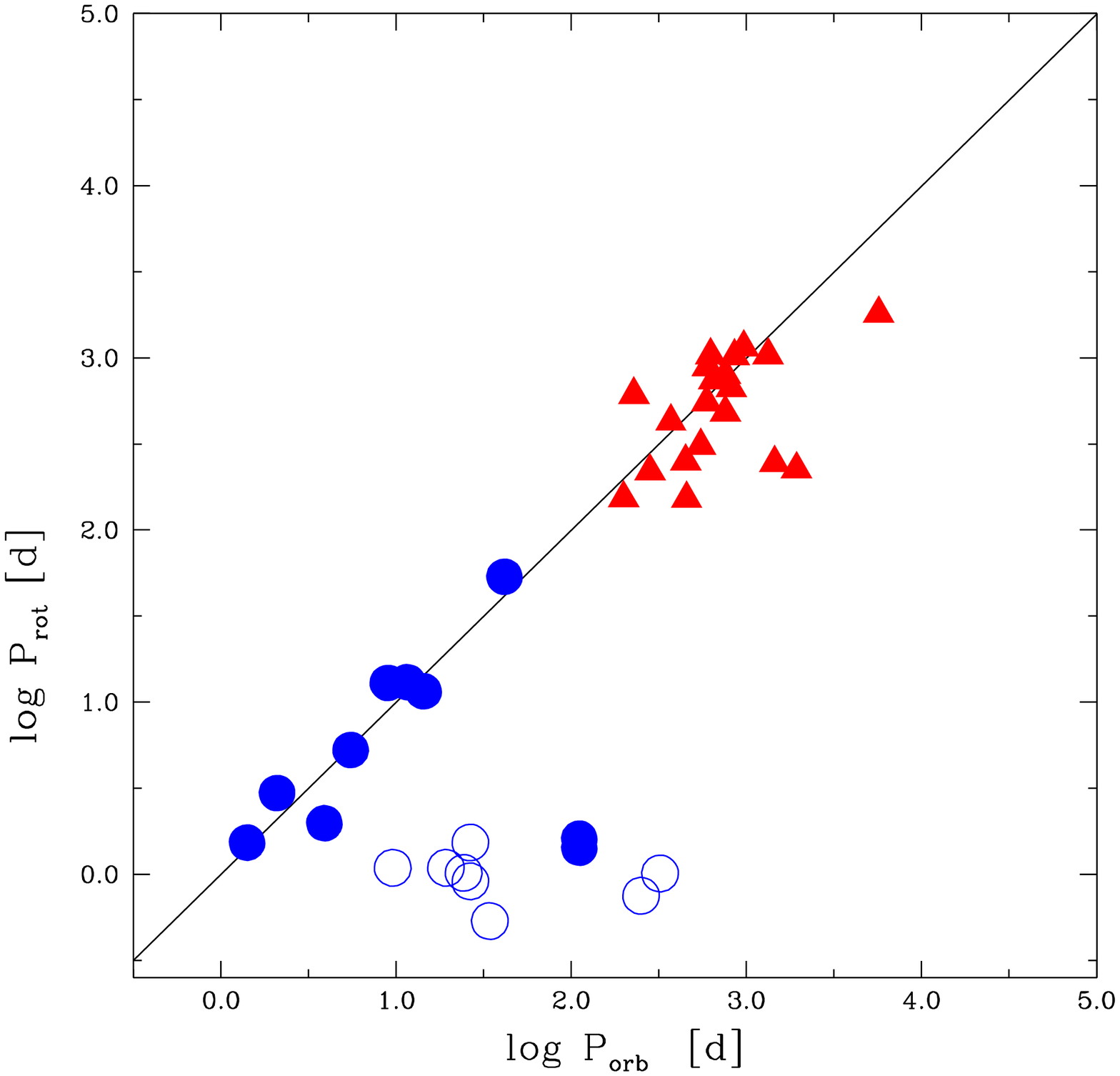}}}
\caption{P$_{rot}$ versus P$_{orb}$ on a logarithmic scale. 
      The blue circles indicates the High-mass X-ray Binaries - filled symbols for giant/supergiant systems and
       open symbols for Be/X-ray binaries. 
      The red triangles indicates the S-type symbiotic stars. }
\label{stoyanov-fig1}
\end{myfigure}

\vspace{-10mm}
\section{Orbital eccentricity of MWC~560}
MWC~560 is a symbiotic star,
which consists of a red giant and a white dwarf   
(Tomov et al. 1990).
The most spectacular features of this object are the collimated ejections 
of matter with velocities of up to $\sim 6000$~\kms\ (Tomov et al. 1992)
and the resemblance of its emission line spectrum to that of 
the low-redshift quasars (Zamanov \& Marziani, 2002).
The jet ejections are along the line of sight and the 
system is seen almost pole-on ($i < 16^\circ$).
This makes it difficult to obtain the orbital eccentricity of the system in a conventional way.

For the system we adopt 
$R_{\rm g} = 140 \pm 7~R_\odot$, $L_{\rm g} \sim 2400~L_\odot$,
 $M_{\rm g} = 1.7~M_\odot$,
$M_{\rm wd}=0.65~M_\odot$, $P_{\rm orb}=1931 \pm 162$~day (Gromadzki et al. 2007), $R_{\rm env}=0.9~R_{\rm g}$ and $M_{\rm env}= 1.0~M_\odot$ (Herwig 2005). 
With the above values of the parameters assumed, we derive the semi-major axis of the orbit 
to be $a \approx 860~R_\odot$.
Using these parameters, we calculate from Eq.~\ref{sync} and Eq.~\ref{circ} the synchronization and circularization time scales:
$\tau_{\rm sync} = 2.6 \times 10^4$~yr and 
$\tau_{\rm circ} = 3.1 \times 10^6$~yr.
The typical lifetime of a symbiotic star is $\tau_{\rm ss} \sim 10^5$~yr (Yungelson et al., 1995). 
From the rate of accretion on the white dwarf, 
$\dot M_{\rm acc} \approx 5 \times 10^{-7}~M_\odot$ (Schmid et al. 2001), we can estimate, that it will take $10^6$~yr 
to accrete $\sim 0.5~M_\odot$ from the envelope of the red giant companion. 
Because the giant also losses mass via stellar wind, we find that the lifetime 
of the symbiotic phase of MWC~560 should be 
$\tau_{\rm ss} \leq 10^6$~yr.
For MWC~560 we have therefore the situation in which $\tau_{\rm ps} < \tau_{\rm syn} < \tau_{\rm ss} < \tau_{\rm circ}$. 
This means that the symbiotic phase is long enough that the tidal forces can 
(pseudo)synchronize the rotation of the red giant. 
On the other hand, the value of $\tau_{\rm circ}$ demonstrates that 
the symbiotic lifetime of MWC~560 is shorter than the circularization time, 
and therefore the orbit can be eccentric.
This is in agreement with the observational evidences found by 
Fekel et al. (2007) that the symbiotic stars with P$_{\rm orb} > 800$~days 
tend to have eccentric orbits. The above implies that in MWC 560, the red giant is probably synchronized,
but the orbit is not circularized.
To determine the orbital eccentricity of  MWC~560, we need 
to calculate $P_{\rm rot}$ for the mass donor.
We analyzed 21 high resolution spectra of MWC~560 and obtained value for \vsi $= 8.2 \pm 1.5$~\kms. 
Using $R_{\rm g} = 140 \pm 7~R_\odot$ and  $\; i=12^\circ-16^\circ$,  
we calculate $P_{\rm rot}=144 - 306$~days. This value 
is less than the orbital period.
MWC~560 should 
be close to synchronization or pseudosynchronization, 
and $P_{\rm rot} =P_{\rm ps}$. 
Using Eq.~\ref{Eq-ps} we can estimate the orbital eccentricity to be 
$e=0.68-0.82$.

\vspace{-5mm}
\section{Conclusions}
\vspace{-4mm}
Using rotational velocity measurements and the theory of synchronization/pseudosynchronization we:

(1) find that the Be/X-ray binaries are far away from 
(pseudo)synchronization.
The tidal force in the Be/X-ray binaries acts as a decelerator 
of the rotation of the mass donors.  The only Be/X-ray binary
which is close to pseudosynchronization is the LSI$+61^0303$. 
The objects containing mass donors of spectral class I 
and III typically have $P_{rot} \sim P_{ps}$ and are
close to (pseudo)synchronization;

(2) demonstrate that the M giants in symbiotic stars rotate faster than the field giants.
Most symbiotics with orbital period less than 1000~d are synchronized;

(3) show that the High-mass X-ray binaries and the S-type symbiotic stars are either on the line of
synchronization or they are under the line. None of the objects in our sample is above the line of synchronization.

(4) calculate that the orbit of the symbiotic star MWC~560 should be highly eccentric, with $e\sim$ 0.7. 

\thanks
We thank the anonymous referee for constructive comments. This work was supported by the OP ``HRD``, ESF and Bulgarian Ministry of Education, Youth and Science under the contract 
BG051PO001-3.3.06-0047.

\end{multicols}
\end{document}